\documentclass[a4paper, 10pt, conference]{IEEEtran}
\IEEEoverridecommandlockouts
\usepackage{cite}
\usepackage{multirow}
\usepackage{amsmath,amssymb,amsfonts}
\usepackage{algorithm}
\usepackage{algpseudocode}
\usepackage{graphicx}
\usepackage{textcomp}
\usepackage{xcolor}
\usepackage{hyperref}
\usepackage{cleveref}
\usepackage{amsmath}

\begin{document}

\title{Bayesian propensity score matching in automotive embedded software engineering}

\author{\IEEEauthorblockN{Yuchu Liu}
\IEEEauthorblockA{
\textit{Volvo Cars}\\
Gothenburg, Sweden \\
yuchu.liu@volvocars.com}
\and
\IEEEauthorblockN{David Issa Mattos}
\IEEEauthorblockA{\textit{Computer Science and Engineering} \\
\textit{Chalmers University of Technology}\\
Gothenburg, Sweden \\
davidis@chalmers.se}
\and
\IEEEauthorblockN{Jan Bosch}
\IEEEauthorblockA{\textit{Computer Science and Engineering} \\
\textit{Chalmers University of Technology}\\
Gothenburg, Sweden \\
jan.bosch@chalmers.se}
\and
\IEEEauthorblockN{Helena Holmstr\"om Olsson}
\IEEEauthorblockA{\textit{Computer Science and Media Technology} \\
\textit{Malm\"o University}\\
Malm\"o, Sweden \\
helena.holmstrom.olsson@mau.se}
\and
\IEEEauthorblockN{Jonn Lantz}
\IEEEauthorblockA{
\textit{Volvo Cars}\\
Gothenburg, Sweden \\
jonn.lantz@volvocars.com}
}

\maketitle

\begin{abstract}
Randomised field experiments, such as A/B testing, have long been the gold standard for evaluating the value that new software brings to customers. However, running randomised field experiments is not always desired, possible or even ethical in the development of automotive embedded software. In the face of such restrictions, we propose the use of the Bayesian propensity score matching technique for causal inference of observational studies in the automotive domain. In this paper, we present a method based on the Bayesian propensity score matching framework, applied in the unique setting of automotive software engineering. This method is used to generate balanced control and treatment groups from an observational online evaluation and estimate causal treatment effects from the software changes, even with limited samples in the treatment group.
We exemplify the method with a proof-of-concept in the automotive domain. In the example, we have a larger control ($N_c=1100$) fleet of cars using the current software and a small treatment fleet ($N_t=38$), in which we introduce a new software variant. 
We demonstrate a scenario that shipping of a new software to all users is restricted, as a result, a fully randomised experiment could not be conducted. 
Therefore, we utilised the Bayesian propensity score matching method with 14 observed covariates as inputs. The results show more balanced groups, suitable for estimating causal treatment effects from the collected observational data. 
We describe the method in detail and share our configuration. Furthermore, we discuss how can such a method be used for online evaluation of new software utilising small groups of samples.
\end{abstract}


\section{Introduction \label{intro}}

Over the past decade, field experiments (or, online experiments) have become a ubiquitous part of software development. Software-as-a-Service (SaaS) companies have long shared success stories of the use of experiments to assess the value software features deliver to users\cite{google2010, Deng2013, Xie2016}. These success stories have led companies beyond the SaaS domain, specifically automotive companies, to show interest and even start running experiments \cite{Giaimo2017, Mattos2018, Mattos2020, Giaimo2020, Liu2021}.
Nevertheless, the automotive domain faces many unique restrictions compared to SaaS companies, such as number of software variants, architecture restrictions, safety-regulation constraints, number of vehicles available for experimentation, driver consent, and the ability to frequently update software in customer vehicles due to limitations such as user and privacy agreements among others \cite{Mattos2020, Giaimo2020}. A combination of these challenges leads to many situations were the design of an online experiment is not: (1) possible, such as in limited samples; (2) desired, such as in safety-critical systems; (3) ethical, such as without explicit consent of the drivers on the scope of the software. 

These restrictions to properly conduct a field experiment require that the research and development organisation to utilise a range of different causal inference techniques to assess the value delivered by the new software. 
In this paper, we propose the use of the Bayesian propensity score matching technique for observational causal inference in the automotive domain. We introduce the BOAT (\textbf{B}ayesian propensity score matching for \textbf{O}vserv\textbf{A}tional \textbf{T}esting) method.
This method is used to generate balanced control and treatment groups from observational data and estimate causal treatment effects from software changes, even with limited samples. 
The BOAT method is based on the propensity score matching framework by Rubin \cite{Rubin1996}.
The propensity score matching framework has been developed and wildly applied in medical science \cite{Rubin2001, Xu2009, Stuart2010, Stuart2010a}, in traffic safety analysis \cite{Li2020, Zhang2021}, in SaaS systems \cite{Xu2016}, and in automotive software for experiment design \cite{Liu2021}.


We demonstrate the BOAT method using a proof-of-concept in the automotive domain. 
We ship a modified software variant to a part of our case company's internal fleet, and compare that to a larger population of vehicles that are equipped with the existing variant of the software. 
As in the automotive sector, the access to update customer vehicles is significantly more limited than data collection.
When we can collect data from more vehicles than we can ship software to, we have skewed sample sizes in the control and treatment group.
Our proof-of-concept is designed to simulate such scenarios.
Therefore, the new software is only download to a limited number of vehicles, these vehicles are driven by the employees as their primary personal cars.
The control group ($N_c=1100$) of cars uses the current software variant and the treatment group ($N_t=38$) utilises the new software.
We collect measurements from vehicle on-board sensors for a continuous period of five months, engineer the input features to BOAT, and perform a matching to produce control/treatment groups with balanced empirical distribution of the features.
Note that, features and covariates refer to the independent variables in statistical models, and we use the two terms interchangeably.

Comparing to the existing literature, this paper provides the following contributions.
First, we describe the theoretical background of Bayesian propensity score matching model. To the best of our knowledge, this is the first time such a model is used in software development publications. 
Second, we discuss the feasibility of such an application in automotive software engineering. 
Third, we share the process of rolling out small-scale observational testing of automotive software.
In combination with the BOAT model, we are able to introduce observational testing of novel software in a fast and more robust manner, and conclude the causal effects of such software change from observational studies. 

The rest of the paper is organised as follows. In Section \ref{background}, background and related work are introduced. 
In Section \ref{bpsm}, we describe the Bayesian propensity score matching theory in detail.
We present the data structure and collection method in Section \ref{method}.
The results are presented in Section \ref{case}.
The discussion and conclusion are presented in Section \ref{discussion} and Section \ref{conclusion} respectively.
\section{Background and related work\label{background}}

To evaluate software changes, companies use randomised field experiment techniques such as A/B testing \cite{google2010, kohavi2009controlled, Deng2013, Xie2016}.
In their experiments, users are randomly split into two large groups and introduced to different variants of the software.
The variants usually include the existing variant (control) and a modified one (treatment).
The assumption of such online experiments is that the sample size is large enough, thus the two or more groups are balanced and directly comparable, and the only difference is the software variant.
When the assumption holds, the experimenters can establish a causal relation in between the software change and the treatment effect through measuring carefully designed metrics.
Online experiments conducted by SaaS companies benefit from their large user base.
Yet unbalanced groups could be produced in those experiments due to high diversity in users \cite{Deng2013, Xie2016}, known confounding factors such as user preferences \cite{Xu2016}, or other unknown confounding factors \cite{Gupta2018}.
These issues are addressed through techniques such as propensity score matching for online quasi-random experiments \cite{Xu2016}, stratified sampling and the CUPED method (Controlled experiment Using Pre-Experiment Data) \cite{Deng2013, Xie2016}.

Conducting large and fully randomised online experiments can be more challenging in the automotive domain \cite{Giaimo2017, Mattos2018, Mattos2020, Giaimo2020}.
First, the available users in the automotive domain are comparably more limited than in SaaS, and most manufactures have a high diversity in their products (e.g., vehicle customised options), which will further reduce the available samples for large online experiments \cite{Mattos2020, Liu2021}.
Second, a vast majority of automotive software are in safety critical systems \cite{Giaimo2017, Giaimo2019}.
Although the risks of safety compromise are minimum since the modified software variants will go through the same release process \cite{Mattos2020}, but even minor disturbances at a scale can be directly translated to the profit lost for commercial vehicles such as trucks and taxis.
Thus, it is undesirable to ship software for safety critical systems to a large portion of the vehicle fleet at once.
Last but not least, software update and data collection requires explicit consent from the vehicle users.

As a result, comparing to large and fully randomised online experiments, automotive software online evaluation is much more feasible to be done in the format of small-scale observational studies, where the new software is only introduced to a small and selective group of vehicles.
An observational study is to be conducted when a randomised experiment is not feasible and a causal relation of treatment and effect is to be established \cite{Cochran1965}.
Therefore, it is critical to present causal inference methods, such as propensity score matching.
With propensity score matching, one can utilise pre-experimental data to design balanced control and treatment groups, as previously demonstrated by \cite{Rubin2001, Xu2009} in the medical sector and \cite{Liu2021} in the automotive domain.
Moreover, propensity score matching has been applied in the field of software engineering, in the efforts of analysing development efficiency \cite{Ramasubbu2009, Tsunoda2017}.

Bayesian propensity score matching (BPSM) is an extension of the traditional framework of propensity score matching, in which the propensity score is estimated through a Bayesian network.
Using Bayesian statistics, one can conjugate the posterior distribution based on a prior, a likelihood and evidence.
In other words, Bayesian statistics allows one to model based on the data and the domain knowledge \cite{Torkar2020}.
Instead of providing only a point estimate of the dependent variable, Bayesian models will return the entire posterior distribution, therefore, quantifying uncertainty.
BPSM, as applied in the filed of traffic research \cite{Li2020}, has shown a higher performance than the frequency approach for small samples.
\section{Bayesian propensity score matching \label{bpsm}}


In this section, we present the theory of Bayesian propensity score matching (BPSM) in detail.
A probabilistic graphical model is used to illustrate the Bayesian logistic regression generative model, and we present the prior, the evidence and the posterior in this Bayesian network.
Finally, we describe different matching strategies and the ones applied in the paper.

Propensity score matching, first introduced by Rosenbaum and Rubin \cite{Rosenbaum1983} in 1983, is a causal inference model for estimating treatment effects from observational studies.
In an observational study, the measured treatment effects could be caused by confounding variables than the treatment itself, thus raise bias in the results.
When the sample sizes are limited, propensity score matching can help us create balanced and comparable groups by matching the control and treatment groups, so that the covariates from both groups form similar empirical distributions.
Propensity score matching can be used to design partitioning of control and treatment groups based on pre-experimental observations \cite{Rubin2001, Xu2009, Stuart2010a, Liu2021}, or used for causal treatment/no-treatment effect analysis of existing observational studies postmortem \cite{Xu2016, Li2020, Zhang2021}.
The most important assumption of propensity score matching is ignorability \cite{Stuart2010}, which implies the unobserved covariates do not influence the target variable, thus ignorable. 
In other words, propensity score matching can only balance covariates that are observed but a fully randomised experiment with a large sample can balance all covariates, observed or not. 

In a two-group observation study with total sample size $N$, the average treatment effect ($ATE$) is defined as the difference of the average expected value of the target variable in the control ($E(z_n|y_n = 0)$) and treatment group ($E(z_n|y_n = 1)$),

\begin{equation}
    ATE = \frac{1}{N} \sum_{n=1}^{N} (E(z_n|y_n=1) - E(z_n|y_n=0))
\end{equation}

Where $y_n \in \{0, 1\}$ is a control ($y_n = 0$) or treatment ($y_n = 1$) indicator for each sample $n = \{1, 2, ..., N\}$. 
For each $n$, we observe a total of $I$ numbers of covariates $x_i$, $x_i = \{x_1, x_2, ... x_I\}$, which are correlated with the target variable $z_n$, denotes as $\textbf{x}_n$ for all samples $N$. 
The $\textbf{x}_n$ is a matrix with dimension $N \times I$. 
Covariates $\textbf{x}_n$ are the confounding variables that would potentially influence the target variable $z_n$.
The potential outcome of the target variable is independent of the treatment assignment given the covariates,

\begin{equation}
    (z_{n, c}, z_{n, t}) \bot y_n | \textbf{x}_n
\end{equation}

The average expected treatment effect becomes conditional to both treatment $y_n$ and the covariates $\textbf{x}_n$,

\begin{equation}
       ATE_{PSM} =
        \frac{1}{N} \sum_{n=1}^{N} (E(z_n|x_n, y=1) - E(z_n|x_n, y=0))
\end{equation}

There are two steps in propensity score matching. The first step is the estimation of propensity score through logistic regression followed by performing matching of samples in the control and treatment groups based on their propensity scores.
In BPSM, the estimation of the propensity score is done through a Bayesian logistic regression, which returns a mean propensity score for each sample and their uncertainties.

    \subsection{Probabilistic graphical model}

\begin{figure}[t]
\centerline{\includegraphics[width=\linewidth]{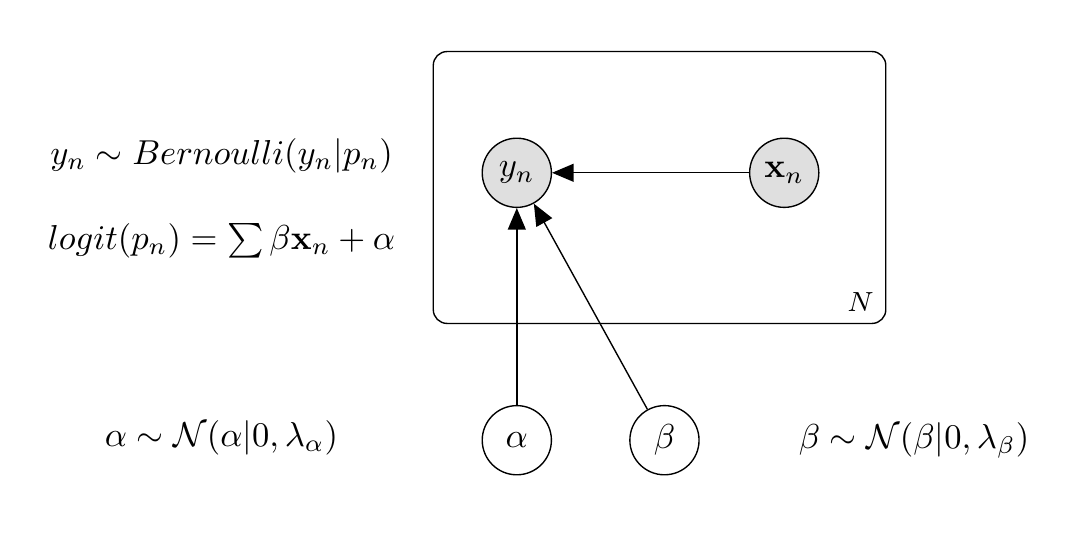}}
\caption{Probably graphical model of a Bayesian logistic regression, with observed input features ($\mathbf{x_n}$), treatment indicator ($y_n$), and latent variables as regression model coefficients ($\alpha, \beta$).}
\label{fig_pgm}
\end{figure}

A probabilistic graphical model for Bayesian network is a directed acyclic graph, in which the shaded nodes represent the observed variables such as features and treatment indicator variable. 
The bright nodes are latent variables.
The directional edges indicate conditional dependencies in between variables, and the unconnected nodes are conditionally independent.
The plate is a representation of the number of observations, i.e., samples.
The first step in BPSM is to estimate the propensity score through a logistic regression.
A probabilistic graphical model for Bayesian logistic regression is shown in Fig. \ref{fig_pgm}.

Consider a non-randomised study where the users in the treatment group are not randomly assigned and there are only a limited number of users in the group. 
We have a total number of samples $N$, in which there are more or equal number of samples in the control group ($N_c$) than the treatment group ($N_t$), $N_c \geq N_t$.
The regression coefficients, $\alpha$ and $\beta$ are latent variables. 
That is they are not observed but inferred from other variables that are observed.
The treatment indicator $y_n$ is binary, and it follows a Bernoulli distribution,

\begin{equation}
    y_n \sim Bernoulli(y_n|p_n)
\end{equation}

where the propensity score $p_n$ is calculated as,
\begin{equation}
    p_n = \frac{e^{\alpha  + \beta \mathbf{x}_{n}}}{1 + e^{\alpha  + \beta \mathbf{x}_{n}}}  
\end{equation}

The regression intercept $\alpha$ has a prior of Gaussian distributions of $0$ mean and a variance of $\lambda_{\alpha}$,
\begin{equation*}
  \alpha \sim \mathcal{N}(\alpha|0, \lambda_{\alpha})
\end{equation*}

similarly, the regression coefficient $\beta$ has a prior of Gaussian distributions of $0$ mean and variance of $\lambda_{\beta}$,
\begin{equation*}
    \beta \sim \mathcal{N}(\beta|0, \lambda_{\beta})
\end{equation*}


Bayesian networks are generative models and to generate the joint probability distribution of the regression model, the generative process is stated as the following Algorithm \ref{bpsm_gen}.


\begin{algorithm}
\caption{Bayesian logistic regression generative process}
\textbf{Inputs: } $\mathbf{x}_n$ covariates, $\lambda_{\alpha}$ prior distribution of $\alpha$, $\lambda_{\beta}$ prior distribution of $\beta$, $y_n$ control/treatment indicator
\begin{algorithmic}[1]
\State Draw $\alpha \sim \mathcal{N}(\alpha|0, \lambda_{\alpha})$
\State Draw $\beta \sim \mathcal{N}(\beta|0, \lambda_{\beta})$

\For{each vector of covariates $x_i$ in $\{x_1, x_2, ..., x_I \}$}
     \State Draw $y_n \sim Bernoulli(y_n|Sigmoid(\alpha + \beta x_n ))$
\EndFor
\end{algorithmic}
\label{bpsm_gen}
\end{algorithm}

By Bayesian Theorem, the posterior distribution of the network is the product of the likelihood and the prior.
In this case, the posterior distribution is a joint probability of $y_n$, $\alpha$, and $\beta$ marginalised over $p(y_n)$, that is,

\begin{equation}
\begin{array}{l}
    p(y_n, \alpha, \beta | \textbf{x}_n, \lambda_{\alpha}, \lambda_{\beta}) \\
        = p(\alpha|\lambda_{\alpha}) \cdot  p(\beta|\lambda_{\beta}) \cdot \prod_{n=1}^{N}p(y_n|\alpha, \beta, x_n)
\end{array}
\end{equation}

In many cases, the exact posterior distribution cannot be solved analytically, but it can be approximated with stochastic (e.g., Markov Chain Monte Carlo) or deterministic (e.g., variational inference) methods.
In this paper, we approximate the posterior distribution through a stochastic method, the No-U-Turn Sampler (NUTS) in Hamiltonian Monte Carlo algorithm.
Using a recursive algorithm, NUTS constructs a set of possible candidate point spans widely across the target distribution \cite{Hoffman2011}. NUTS stops automatically if it retraced its steps, hence the name "No-U-Turn".
We set up a NUTS sampler with a single chain, 3000 samples, and 200 warm-up samples were discarded.
We include the model setup of the Bayesian logistic regression, the inference solver, and the trace plots in the online appendix.

    \subsection{Matching}

After inferring the propensity score from the Bayesian logistic regression model, the second step of BPSM is to match the control and treatment pairs based on their propensity score distances.
The objective of the matching is to form balanced control and treatment groups, that is, minimising the propensity score distance.

The propensity score distance ($\delta p_n$) is defined as the absolute difference of the propensity score in the control and treatment group,

\begin{equation}
    \delta p_n = |p_{n, \tau=0} - p_{n, \tau=1}|
\end{equation}

There are a few different methods for matching, such as calliper matching \cite{Austin2011}, 1:1, or n:1, nearest neighbour matching \cite{Stuart2010a}, and full matching \cite{Xu2009, Hansen2004}.
Matching can be done with or without replacement.
When matching with replacement, one sample in the control group can be matched with multiple samples in the treatment group, and vice versa. 
Full matching method matches with replacement, such as the optimal full matching algorithm \cite{Hansen2004}.

There are matching methods that do not allow sample replacement, and by using such matching methods, the matched control and treatment groups will have the same number of samples.
Commonly applied methods include calliper matching, where the highest permitted calliper for $\delta p_n$ is predetermined, and control and treatment pairs will be matched based on this calliper.
Calliper matching is computationally cheap and intuitive \cite{Austin2011}, however, it could result in a reduction in number of samples in the treatment groups if the propensity score distances fall out of the predefined calliper.
Sometimes it could be less ideal to reduce sample numbers in the treatment group, as they are usually considered as expensive samples especially in the embedded domain.
Another matching method, 1:1 nearest neighbour matching, selects one controlled sample to match one treated sample with the smallest distance $\delta p_n$ \cite{Stuart2010a}.
Using 1:1 nearest neighbour matching, no samples from the treatment group will be reduced and the sample reduction will only happen in the control group.
In this paper, we apply both calliper matching and 1:1 nearest neighbour matching.
As our objective is to find samples in the control group to be compared with the treated samples.
\section{Input data \label{method}}

\begin{table*}[t]
\caption{Descriptive statistics of the target variable and covariates, and a description of how the variables are computed. Each variable is aggregated to the vehicle level and max-min scaled.}
\centering
\begin{tabular}{lllll}
\hline
\textbf{Variables} & \textbf{Variable description} & \textbf{Group} & \textbf{Mean} & \textbf{Std.} \\ \hline\hline
\textbf{Target variable} &  &  &  &  \\
\multirow{2}{*}{Fuel consumption {[}g/km{]}} & \multirow{2}{*}{total fuel injected in engine / total distance} & Control & 0.391 & 0.155 \\
 &  & Treatment & 0.354 & 0.123 \\ \hline
\textbf{Covariates} &  &  &  &  \\
\multirow{2}{*}{Share of trip start at a high state-of-charge} & \multirow{2}{*}{number of trip where soc\_start \textgreater 80\%} & Control & 0.356 & 0.177 \\
 &  & Treatment & 0.397 & 0.178 \\
\multirow{2}{*}{Share of trip end at a low state-of-charge} & \multirow{2}{*}{number of trip where soc\_end \textless 21\%} & Control & 0.258 & 0.155 \\
 &  & Treatment & 0.120 & 0.150 \\
\multirow{2}{*}{Number of trips made on weekdays} & \multirow{2}{*}{number of trips taken place during weekdays} & Control & 0.290 & 0.167 \\
 &  & Treatment & 0.225 & 0.153 \\
\multirow{2}{*}{Number of trips made on weekends} & \multirow{2}{*}{number of trips taken place during weekends} & Control & 0.269 & 0.173 \\
 &  & Treatment & 0.190 & 0.154 \\
\multirow{2}{*}{Average trip distance {[}km{]}} & \multirow{2}{*}{total trip distance / total number of trips} & Control & 0.301 & 0.132 \\
 &  & Treatment & 0.343 & 0.124 \\
\multirow{2}{*}{Maximum trip distance {[}km{]}} & \multirow{2}{*}{longest trip occurred during the observation period} & Control & 0.278 & 0.193 \\
 &  & Treatment & 0.240 & 0.186 \\
\multirow{2}{*}{Average trip speed {[}km/h{]}} & \multirow{2}{*}{total trip distance / total trip duration} & Control & 0.575 & 0.110 \\
 &  & Treatment & 0.624 & 0.117 \\
\multirow{2}{*}{Maximum trip speed {[}km/h{]}} & \multirow{2}{*}{highest trip speed occurred during the observation period} & Control & 0.637 & 0.125 \\
 &  & Treatment & 0.640 & 0.135 \\
\multirow{2}{*}{Share of distance on "hybrid"} & \multirow{2}{*}{distance driven when vehicle is in hybrid mode / total distance} & Control & 0.956 & 0.103 \\
 &  & Treatment & 0.987 & 0.033 \\
\multirow{2}{*}{Share of trips with a trailer attached} & \multirow{2}{*}{numbers of trip with trailer attached / total number of trips} & Control & 0.034 & 0.081 \\
 &  & Treatment & 0.034 & 0.075 \\
\multirow{2}{*}{Average number of engine starts in a trip} & \multirow{2}{*}{total occurrence of combustion engine RPM \textgreater 500 / total number of trips} & Control & 0.169 & 0.112 \\
 &  & Treatment & 0.176 & 0.176 \\
\multirow{2}{*}{Average ambient temperature {[}$^\circ C${]}} & \multirow{2}{*}{average temperature measured at car during the observation period} & Control & 0.371 & 0.084 \\
 &  & Treatment & 0.372 & 0.071 \\
\multirow{2}{*}{Minimum ambient temperature {[}$^\circ C${]}} & \multirow{2}{*}{minimum temperature measure at car during the observation period} & Control & 0.497 & 0.135 \\
 &  & Treatment & 0.563 & 0.150 \\
\multirow{2}{*}{Maximum ambient temperature {[}$^\circ C${]}} & \multirow{2}{*}{maximum temperature measure at car during the observation period} & Control & 0.388 & 0.101 \\
 &  & Treatment & 0.374 & 0.099 \\ \hline
\end{tabular}
\label{table_data}
\end{table*}

In this section, we will present the data used for the BPSM model, the collection methods, and how each data feature is engineered from the measurements.
The data collection is done from the 26th of October 2020 to the 22nd of March 2021.
The measurements are collected from two specific vehicle models.

The measurement of the vehicles are done through on-board sensors for vehicle control, calibration, and diagnostics.
We select low level signals for their robustness and reliability. 
The measurements are done during each drive cycle of the vehicle in a time series format at ten hertz frequency, marked by an arbitrary vehicle ID that cannot be decoded to identify the user nor the vehicle, and a  drive cycle ID.
Drive cycle refers to the events in between each vehicle key-on and key-off, and a trip for the user could consist multiple drive cycles.
The measurements are sent to a central server of the case company through a telecommunications module in the vehicle, no physical access is needed to obtain the data.

To further increase the robustness of the measurements collected, we read the values from two or more sensor signals for each measurement.
We intend to exclude drive cycles in which multiple signals yield drastically difference values. 
However, we found that the same measurement calculated from different signals only differ by a decimal point on the drive cycle level.
For example, the drive cycle distance can be calculated through integrating the instantaneous vehicle velocity, or through a wheel speed sensor that measures the angular velocity of the wheels.
These two signals differ by 0.0227 kilometres for 75 percentile of drive cycles.
Moreover, we measure values in base units. E.g., fuel consumption is measured in grams, because the commonly used unit litre is a secondary value dependent on the pressure and temperature.
Furthermore, we do not include entries if they have any of the properties listed below.
After postprocessing, we have in total 421,881 drive cycles made by 1138 vehicles of which 38 are in the treatment group.

\begin{itemize}
    \item Drive cycle is made by vehicles with odometer distance less than 100 kilometres, i.e., brand new vehicles.
    \item Drive cycle average speed is greater than 200 km/h.
    \item Drive cycle total distance is less than 0.5 kilometres.
    \item Drive cycle total duration is less than one minute.
\end{itemize}

    \subsection{Data structure}

The input data features to BPSM model is produced from the time series values collected from the vehicles. 
Note that due to our confidentiality agreement with the company, the input data will not be shared nor shown without scaling in this paper.
First, we aggregate each measurement per drive cycle through multiple vehicle signals.
Since multiple signals do not return a different outcome beyond a decimal point, we select one value to keep.
After this step, we produce a dataframe that compresses of one drive cycle per vehicle per row.

Second, we calculate the features based on all trips per vehicle and produce 14 input features and one target variable to BPSM.
The input features are stored in a matrix with dimension $1138 \times 14$, which corresponds to 1138 vehicles and 14 features.
The features and how they are calculated are presented in Table \ref{table_data} along with the descriptive statistics.
Each feature is scaled with their perspective minimum and maximum values.

\section{Results \label{case}}

In this section, the results of the Bayesian propensity score matching are presented.
We show the propensity score computed from Bayesian logistic regression, along with matched groups from two different matching methods.  
Finally, we present the process of Bayesian propensity score matching for Observational Testing for evaluating software online.

    \subsection{Bayesian propensity score}

Following the generative process described in \ref{bpsm_gen}, a Bayesian logistic regression model is implemented in Pyro \cite{bingham2019pyro}.
The model takes covariates $\mathbf{x}_n$ and the prior distributions of $\alpha$ and $\beta$ as inputs, and returns tensors of posterior distributions of $\alpha$ and $\beta$.
A NUTS sampler in Hamiltonian Monte Carlo is used to infer the posterior distribution. 
We set up the sampler with a single chain, 3,000 samples, and 200 burn-in.
Moreover, we apply a variational inference method to triangulate the results. 
The Brooks-Gelman-Rubin convergence criteria of $\hat{R} < 1.1$ is met, at $\hat{R} = 1.0003$.
The variational inference uses a multivariate normal distribution as a guide. We define 40,000 steps for optimisation and the solver reaches a stable solution after the first 10,000 steps.
Two inference methods return similar posterior distributions and point estimates.
We show both methods in the online appendix attached. 
However, we will only focus on reporting the inference results from the NUTS sampler in this section.

\begin{figure}[t]
\centerline{\includegraphics[width=0.5\textwidth]{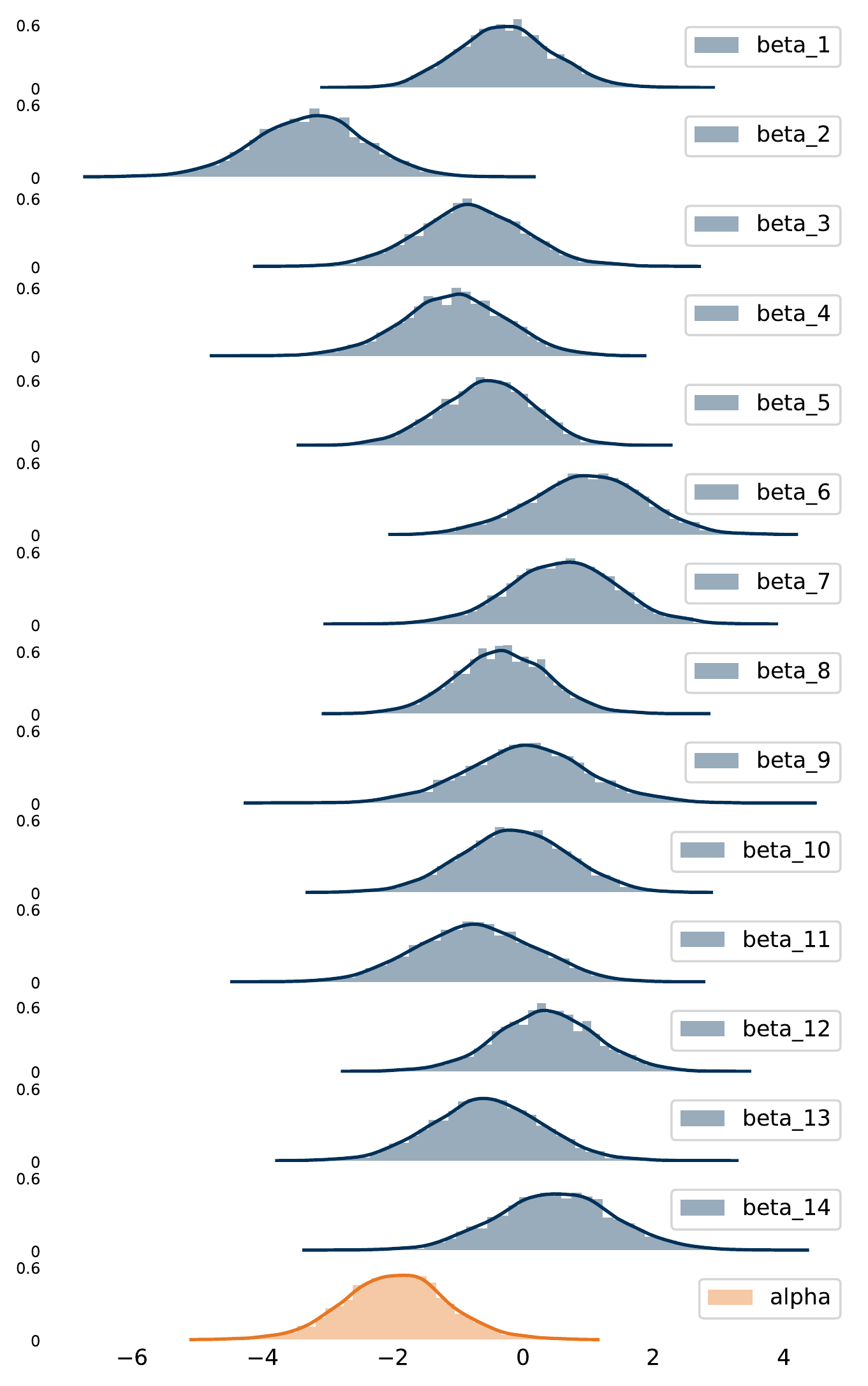}}
\caption{Posterior distributions of the Bayesian logistic regression coefficient $\beta = \{\beta_1, \beta_2, ..., \beta_14\}$, and intercept $\alpha$.}
\label{fig_beta}
\end{figure}

Each regression coefficient $\beta$ has a prior of $\beta \sim \mathcal{N}(0, \lambda_{\alpha} = 1)$, Gaussian distribution, and the regression intercept follows the prior distribution $\alpha \sim \mathcal{N}(0, \lambda_{\beta} = 1)$.
Combining the priors, the posterior, $p(y_n, \alpha, \beta | \mathbf{x}_n, \lambda_{\alpha}, \lambda_{\beta})$, is inferred from the observations $\mathbf{x}_n$ and the evidence $y_n$.
We illustrate the total posterior distributions of all $\alpha$ and $\beta$ in Fig. \ref{fig_beta}.
From the posterior distributions, we can fit the logistic regression from a point estimate that is the mean value of the intercept $\hat{\alpha}$ and the coefficients $\hat{\beta}$, and the uncertainty of the model is quantified from the total posterior distribution. 

\begin{figure*}[ht]
\centerline{\includegraphics[width=\textwidth]{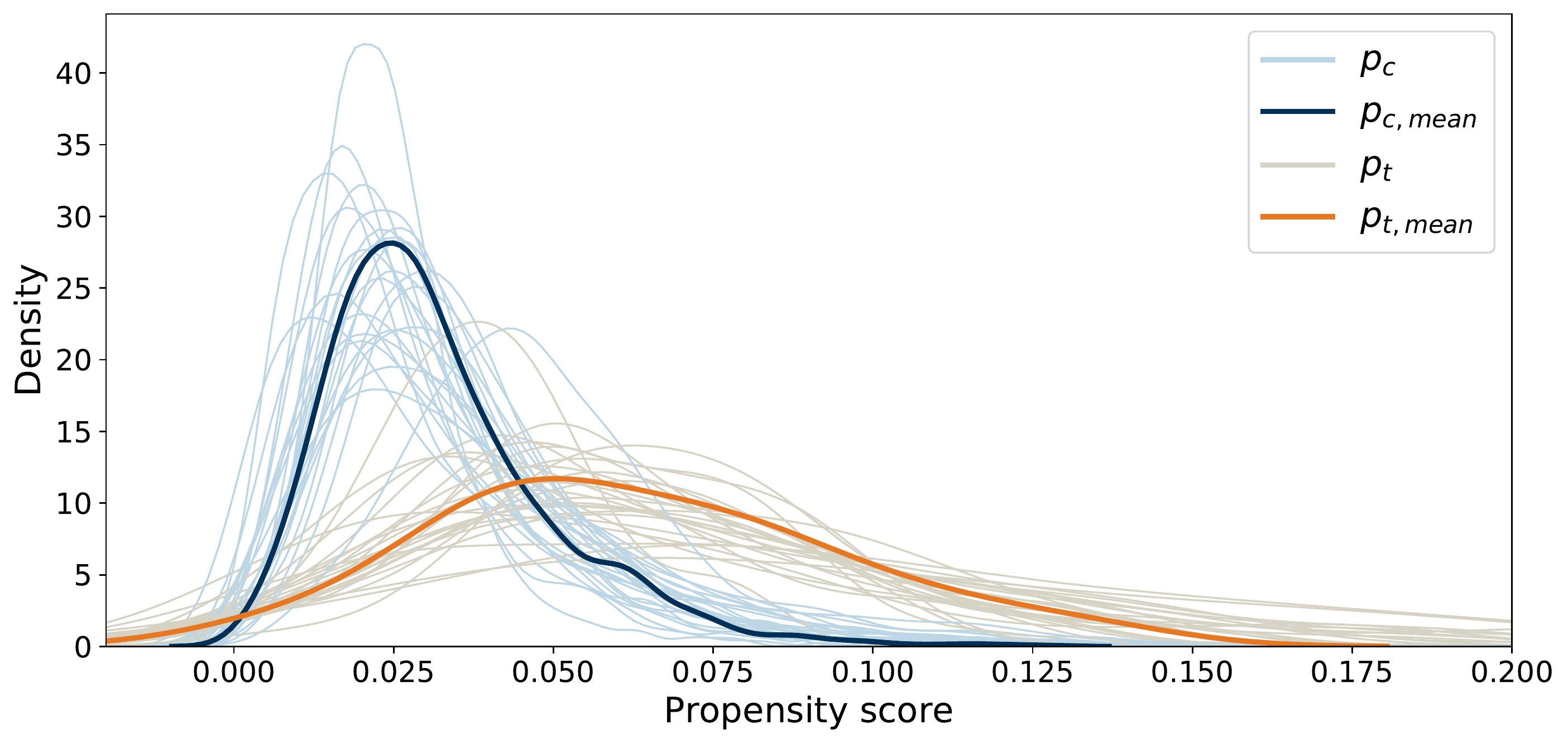}}
\caption{Kernel density distribution of the propensity scores of the control ($p_c$) and treatment ($p_t$) groups calculated on the mean of posterior distributions, and twenty-five values randomly sampled from the posterior distributions representing uncertainties.}
\label{fig_bpsm}
\end{figure*}

The propensity scores for the control group ($p_c$) and the treatment group ($p_t$) are estimated as $e^{\hat{\alpha} + \hat{\beta} \mathbf{x}_{n}} / 1 + e^{\hat{\alpha}  + \hat{\beta} \mathbf{x}_{n}}$.
Without matching, the mean propensity score in the control and treatment group is 0.0319 and 0.0633 respectively. 
In each group, the standard deviation of the propensity score is 0.0175 and 0.0309.
In Figure \ref{fig_bpsm}, we show the kernel density distributions of the propensity scores in the control and treatment groups before performing the matching.
Since the entire posterior distribution is available, we illustrate the uncertainty on the propensity scores by randomly drawing 25 samples from the posterior distributions of the intercept and coefficients, and computing the propensity scores from the drawn $\alpha$ and $\beta$.

    \subsection{Matched A/B groups}

\begin{figure*}[ht]
\centerline{\includegraphics[width=\textwidth]{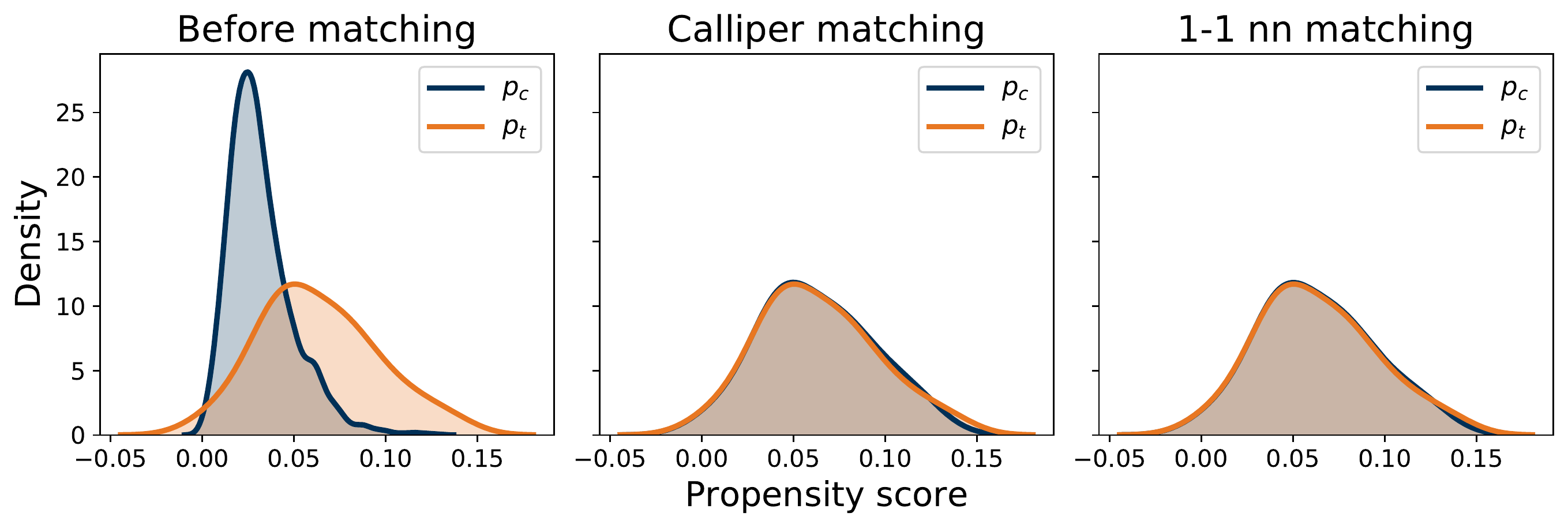}}
\caption{Kernel density distributions of the Bayesian propensity scores for the control ($p_c$) and treatment ($p_t$) group, when, no matching was done, matched with a caliper at 0.05, and matched with 1-1 nearest neighbour.}
\label{fig_matched}
\end{figure*}

After the Bayesian propensity scores are estimated, the second step it to match pairs from the control and treatment group to minimise the propensity score distance $\delta p_n$.
A kernel density plot of the propensity score distribution before and after matching can be found in Figure \ref{fig_matched}.

\begin{table}[t]
\caption{Propensity score in control and treatment groups, before and after matching is applied.}
\centering
\begin{tabular}{llll}\hline
Propensity score &  &  &  \\ 
 & Groups & Mean & Std. \\ \hline\hline
\multirow{2}{*}{Before matching} & Control & 0.0319 & 0.0175 \\
 & Treatment & 0.0633 & 0.0309 \\
\multirow{2}{*}{Calliper matching (calliper = 0.05)} & Control & 0.0626 & 0.0300 \\
 & Treatment & 0.0633 & 0.0309 \\
\multirow{2}{*}{1-1 nearest neighbour matching} & Control & 0.0627 & 0.0302 \\
 & Treatment & 0.0633 & 0.0309 \\ \hline
\end{tabular}
\label{tabel_scores}
\end{table}

Two matching methods are used, both methods find matches without replacement.
The first matching method is a calliper matching, in which a maximum propensity distance is specified and matched pairs are produced if a treated sample has its corresponding pair in the control group.
The number of samples in the control and treatment group is largely skewed in this dataset, using a calliper of 0.05, every treated sample returned a matched controlled sample.
However, if there are too little controlled samples or if the calliper is determined to be too small, calliper matching could return no match for the treated samples.
After calliper matching, the mean propensity score in the control group is 0.0626.

The second matching method is a 1-1 nearest neighbour match.
Similarly to calliper matching, 1-1 nearest neighbour match will return one-to-one matched pairs, but it does so without a specified range of propensity score distance.
For each treated sample, the algorithm k-nearest neighbours searches for one closest neighbour from the control samples. 
The mean propensity score in the control group becomes 0.0627 after matching.
We show the mean and standard deviations of the control and treatment propensity score in Table \ref{tabel_scores}.
On this dataset, both matching methods return similar outcome.
Both methods find corresponding controlled samples for the treated samples.
The average propensity score distance between the control and treatment group is 0.000757 and 0.000608 for the calliper matching and 1-1 nearest neighbour matching respectively.
The covariates balance is assessed by comparing the empirical distribution of covariates in the control and treatment group.
With a calliper matching, we found an average of 4.1 \% reduction in the covariates variance compared to unmatched groups.

    \subsection{Treatment effect}

The treatment, i.e., the new software, is expected to reduce the target variable fuel consumption. 
However, a number of other covariates could influence fuel consumption, such as temperature, trip frequency, trip distance, average speed, and etc.
When the control and treatment group is not partitioned at random and with a large population, it is impossible to conclude a causal effect from the software change even a treatment effect is observed.
Note that this focus of this study is not the actual software performance, thus, the results from this subsection serve as a demonstration.

The treatment effect is analysed using both the calliper matched and 1-1 nearest neighbour matched groups.
The average treatment effect is calculated as the mean difference of the target variable between the control and treatment groups, and all 1138 measured values are min-max scaled.
The average target variable is 0.379 and 0.391 for the control group when matched with calliper and 1-1 nearest neighbour, respectively.  
The average target variable is 0.355 in the treatment group.
The average treatment effect is -0.024 and -0.036 for the control group when matched with calliper and 1-1 nearest neighbour, respectively. 

    \subsection{Bayesian propensity score matching for observational test}

\begin{figure}[t]
\centerline{\includegraphics[width=0.5\textwidth]{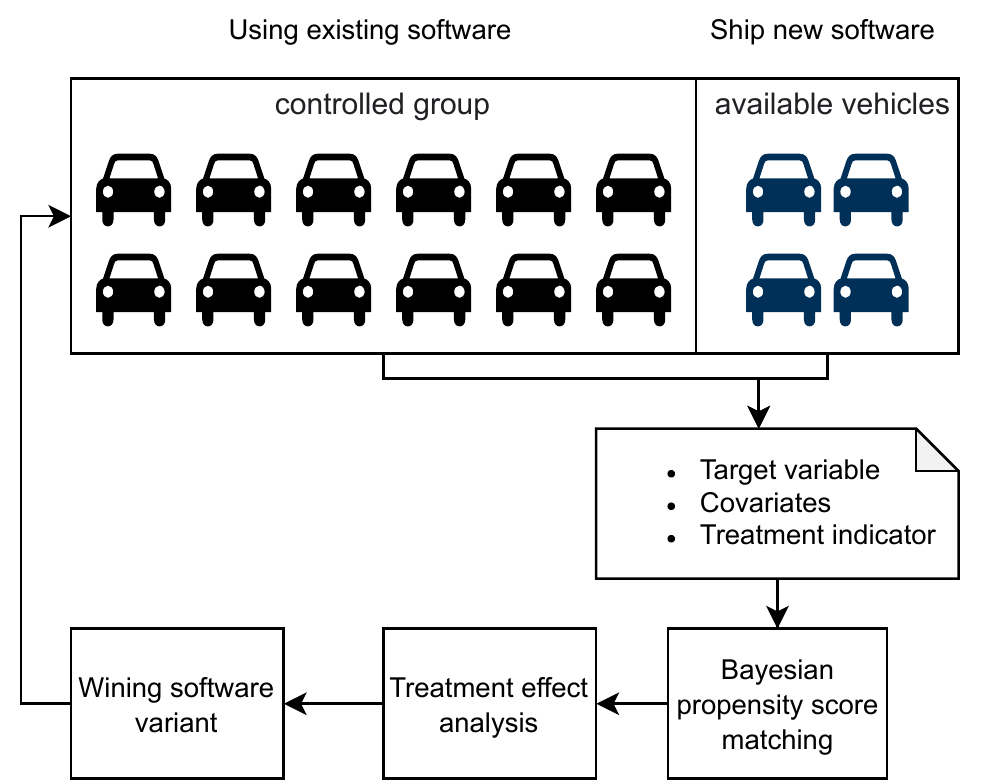}}
\caption{Online software evaluation with limited sample sizes, by utilising Bayesian propensity score matching.}
\label{fig_boat}
\end{figure}

In this subsection, we will describe in detail the process of utilising BPSM for evaluating software online with limited samples, as can be seen in Figure. \ref{fig_boat}.
An observational testing, different from large randomised experiment because the partitioning of control and treatment group is not done at random, and such software online testing is usually done with a very limited sample group.
Furthermore, unlike a pure observational study where no intervention is applied, we introduce treatment to a small cohort.
As discussed previously, with the limitations of automotive embedded software, small-scale observational testing is often the only option in this domain.
To utilise this model for evaluating software online when samples are limited and non-randomised, we recommend the following process.

    \subsubsection{Select treatment software and target variable}

The target variable, that is, the metric of the software evaluation, should reflect the customer and business value the software is aiming to deliver.
The target variable should be measurable. 
Moreover, in rare occasions, some software does not need online evaluation as this additional activity does not add more value to the product. 

    \subsubsection{Determine covariates according to treatment} 

To produce balanced control and treatment groups, covariate selection is important as propensity score matching can only balance the variables that are included in the model. 
The decision on what covariates to be included shall be made both quantitatively and qualitatively. 
The optimal covariates should correlate to the target variable but not the treatment \cite{Brookhart2006}.
The strict statistical correlation between the covariates and the target variable is only part of the inclusion criteria. 
The qualitative domain knowledge of the software and its effect should be taken into consideration, especially when a new software is being evaluated and no high quality user data is available. 

    \subsubsection{Eligible users for the control and treatment group}

The treatment software will be shipped to a subset of users, often to users who have special user agreements in place.
When selecting eligible users for the control group, one needs to make sure their existing software is comparable to the treatment software.
The only systematic difference between the control and treatment group should be the applied treatment, else one could encounter confounding treatment effects from multiple software changes, i.e., a factorial treatment.
In our study, we mitigate this issue by reading the software part number from all effected control units, and only include vehicles with the same part number in the control group.
Additionally, in automotive, some software behaviours are heavily influenced by the devices and their operating locations.
One can include users who drive a certain type of vehicle models in a given country, or include vehicle metadata as categorical variables in the covariates input to BPSM.
    
    \subsubsection{Data collection}

After the sample groups have be determined and the new software is shipped, data collection for both groups starts simultaneously.
The control and treatment vehicles are running in parallel so that seasonality effects can be mitigated. 
The required data collection infrastructure is already in place for our study. 
It is an important enabler for online evaluation of software and we recommend companies to implement data collection capabilities before running any online experiments.
Additionally, a level of understanding of the physical machinery is required when collecting data from embedded systems.
We suggest a cross-disciplinary approach when building such data pipelines for vehicle embedded systems.

    \subsubsection{Run Bayesian propensity score matching}

When the online evaluation has been made and data collected, the Bayesian propensity score matching can be done. 
The propensity score is computed from all the covariates and the treatment indicator in a Bayesian logistic regression, and matching is done accordingly to the matching method and the propensity score.
We implement the Bayesian logistic regression in Python Pyro, as can be found in the online appendix.
The two simple matching methods used in the paper are implemented together with the case company.
The code for the matching algorithms cannot be shared due to our confidentiality agreement, however, some matching algorithms are publicly available in R.
Such as package optmatch\footnote{\href{https://github.com/markmfredrickson/optmatch}{github.com/markmfredrickson/optmatch}} by \cite{Hansen2006}.

    \subsubsection{Assess group balance and analyse treatment effect} 

After the control and treatment groups have been matched with their prospective propensity scores, an assessment of covariates balance should be done.
The covariate balance can be accessed through the absolute standardised mean difference, which compares the absolute difference in means per unit of standard deviation.
Moreover, the mean and variance of each covariate in the control and treatment group should be compared. 
Rosenbaum and Rubin \cite{Rosenbaum1984} suggest an iterative process of diagnostics where additional covariates should be added after an assessment of the groups balance returned from the initial propensity score model.
The average effect is analysed by computing the average difference of the target variable between the matched control and treatment groups.
The new software variant should be introduced if a treatment effect is detected and indicates an improvement.

\section{Discussion \label{discussion}}

In this section, we present the threats to validity in our research and we discuss the generalisability of the Bayesian propensity score matching for observational testing model.
Moreover, we share some known limitation to the Bayesian propensity score matching model, and what the limitations entail when BPSM is applied in online software evaluation in the automotive domain.

\subsection{Threats to validity}
The threats to validity of our research approach are presented in this subsection.
In this paper, we present a proof-of-concept conducted with our case company on a software which optimises energy consumption of hybrid vehicles.
In the treatment group with 38 vehicles, the vehicles are leased to company employees as their company cars and the users have explicit user agreements for participating such tests.
The introduction of the new software variant is made aware to the users, however, we do not disclose the details of the software to them.
Moreover, both the existing and new variants of the software are developed by the case company and we made no inputs to the software itself.

The set of signals measured are predetermined prior to our study, the development teams measure around 500 signals from vehicles, and our data features are engineered from a selected numbers of signals.
We recognise that this means there is a slight risk, some confounding factors might not have been observed in the first place, and their effects on the target variable are unknown to us.
In this study, no special action is taken to mitigate this risk as unobserved and unknown confounding factors should be considered as an inherent limitation of the propensity score matching model.

This study is done on one automotive manufacture and one software.
We accept this limitation to this approach, as the results and conclusions might not be applicable to all software developed by the same company or generalisable to the automotive domain.

First, the piece of software studied does not directly interact with users. 
There is no graphical interface, nor does it require user manual input.
We have not explored how the propensity score matching model reacts to stochastic inputs such as user preferences. 
However, as reported by \cite{Xu2016} from LinkedIn, propensity score matching model is used to support online user-facing software evaluations and shows promising improvements when the user groups are non-randomised.
We foresee similar non-randomised user groups with preexisting preferences in the automotive setting.
Furthermore, we demonstrate the BOAT method using quantitative data measured in the newest vehicles. 
We argue that such quantitative data is rather independent from the vehicle manufacture. Last but not least, we acknowledge companies within the automotive domain could follow difference processes for software development.
Our proposed method that utilises small and non-randomised users for software online evaluation offers a high level of flexibility, and aligns with the core values of the agile methodology that many automotive companies have adopted \cite{Eliasson2014, Mattos2018, Mattos2020}.
As agility is responsiveness to change \cite{Gren2020}, thus, we are optimistic of the value of enabling online software evaluation with small samples in a fast, safe, and ethical manner while maintaining causality.

\subsection{Limitations}

In this subsection, we discuss the limitation of the Bayesian propensity score matching for observational testing method.
First, unlike a fully randomised control and treatment group split which can balance all covariates, propensity score matching can only balance the covariates that are observed.
To make sure the ignorability assumption holds, including the correct covariates is important.
But, when the software is new and there is limited usage data, it can be difficult to have comprehensive knowledge of which covariates should be included in the model.
In this case, an iterative approach can be applied, in which covariates can be added or removed depending on the group balance \cite{Rosenbaum1984}.

Second, Bayesian inference is an expensive method in terms of modelling efforts and computational resources. 
The expense can be justifiable since Bayesian models are flexible as they allow prior input, and they are comprehensive as they return the entire posterior distribution instead of a point estimate which can provide values in post-modelling analysis.
Bayesian propensity score matching returns better results when sample sizes are small but does not show significant improvement as the sample sizes grow \cite{Li2020}.
Therefore, whether to use Bayesian propensity score matching or regular propensity score matching should be a decision made based on the sample size, and a trade-off between computational expense and result improvement.
Finally, we have addressed a scenario where two software versions are to be compared with propensity score matching in this paper, but in practice, multiple candidate software might need to be evaluated.
Propensity score matching can be used for multilevel treatment effect modelling, however less straightforward, as reported by \cite{Imbens2000}. 

%
\section{Conclusion \label{conclusion}}

Online software evaluation is gaining attention in the automotive domain, but large-scale randomised experiments are not always an option with limitations in this industry such as safety and ethics.
In this paper, we present an alternative method to randomise experiments so that online evaluation can be done on small sample groups, enabled with Bayesian propensity score matching model.
This is the first paper to document such a model applied in automotive software engineering.

We describe the theory of Bayesian propensity score matching in detail and demonstrate the model with a proof-of-concept from an automotive company.
In the study, we introduce a new software to a treatment group of 38 vehicles and the control group of 1100 vehicles use the existing software. 
The vehicles in the treatment group are leased to company employees.
We observe both groups for a continuous five month period, during which we collected data from over 400,000 trips.
Data collection is done through the vehicle sensors, and we produce 14 input features to the Bayesian propensity score matching model.
Two matching methods were used, calliper matching and nearest neighbour matching. 
They produce similar results on our dataset and reduce the variance of the covariates by an average of 4.1\%.
Finally, we present the software engineering process of utilising Bayesian propensity score matching for evaluating new functions before shipping them to a larger group of users.
This working method can be complimentary to agile methodologies to enable responsiveness to change and to allow development teams making data-driven decisions.

In our future work in the domain of automotive software online evaluation, we plan to continue to explore and apply different causal inference models.
We see a potential in statistical models which enable online evaluations with limited sample size.
Additionally, we plan to evaluate more automotive software, user-facing functions included, using causal inference methods and develop toolsets for modelling and analysis.

\section*{Online appendix}

We attached an online appendix for the Bayesian logistic regression model.
The online appendix can be found as a Jupyter Notebook via the following link: \href{https://github.com/yuchueliu/BPSM}{github.com/yuchueliu/BPSM}.

\section*{Acknowledgement}
This work is supported by Volvo Cars, by the Swedish Strategic vehicle research and innovation programme (FFI), by the Wallenberg AI Autonomous Systems and Software Program (WASP) funded by the Knut and Alice Wallenberg Fundation and by the Software Center.

\bibliographystyle{IEEEtran}
\bibliography{IEEEabrv, ref.bib}


\end{document}